\documentclass[twocolumn]{aastex63}
\pdfoutput=1
\usepackage{graphicx}
\usepackage{enumerate}
\usepackage{amssymb, amsmath, amsfonts}
\usepackage{gensymb}
\usepackage{mathtools}
\usepackage{physics}
\usepackage{natbib}
\usepackage{hyperref}
\usepackage{color}
\usepackage{ulem}
\usepackage{soul}
\usepackage{url}
\usepackage{xspace}

\newcommand{\codename}{{\tt astrofix}\xspace}
\newcommand{\changes}[1]{{#1}}

\begin{document}

\title{Cleaning Images with Gaussian Process Regression}

\author[0000-0003-4079-2447]{Hengyue Zhang}
\affiliation{Department of Physics, University of California, Santa Barbara, Santa Barbara, CA 93106, USA}

\author[0000-0003-2630-8073]{Timothy D.~Brandt}
\affiliation{Department of Physics, University of California, Santa Barbara, Santa Barbara, CA 93106, USA}

\begin{abstract}
Many approaches to astronomical data reduction and analysis cannot tolerate missing data: corrupted pixels must first have their values imputed. This paper presents \codename, a robust and flexible image imputation algorithm based on Gaussian Process Regression (GPR). Through an optimization process, \codename chooses and applies a different interpolation kernel to each image, using a training set extracted automatically from that image. It naturally handles clusters of bad pixels and image edges and adapts to various instruments and image types. For bright pixels, the mean absolute error of \codename is several times smaller than that of median replacement and interpolation by a Gaussian kernel. 
We demonstrate good performance on both imaging and spectroscopic data, including the SBIG 6303 0.4m telescope and the FLOYDS spectrograph of Las Cumbres Observatory and the CHARIS integral-field spectrograph on the Subaru Telescope. 
\end{abstract}


\section{Introduction} \label{sec:intro}
The history of data artifacts is as long as the history of observational astronomy. Artifacts such as dead pixels, hot pixels, and cosmic ray hits are common in astronomical images. They at best render the pixels' data unusable while at worst disable the entire image in downstream approaches. 

In dealing with missing pixels, some astronomical procedures simply ignore them while others require imputing their values first. Optimal extraction of spectra \citep{Horne_1986} and Point Spread Function (PSF) photometry ignore missing data, while box spectral extraction and aperture photometry do not. Aperture photometry and box extraction have the advantage of requiring little knowledge about the PSF or line-spread function (LSF).  For this reason, aperture photometry has been used for ultra-precise {\it Kepler} photometry \citep{Jenkins+Caldwell+Chandrasekaran+etal_2010}. Box extraction is currently standard for the SPHERE \citep{Claudi_2008} and GPI \citep{Macintosh_2008} integral-field spectrographs.

In general, correcting the corrupted data in an image involves two steps: identifying what they are and imputing their values. Existing algorithms have emphasized bad pixel identification and rejection. For example, there are well-developed packages that detect cosmic rays (CRs) by comparing multiple exposures \citep[e.g.][]{Windhorst_1994}. When multiple exposures are not available, \cite{Rhoads_2000} rejects CRs by applying a PSF filter, \cite{vanDokkum_2001} by Laplacian edge detection ({\tt LACosmic}), and \cite{Pych_2004} by iterative histogram analysis. Among the above methods, {\tt LACosmic} offers the best performance \citep{Farage_2005}. Approaches based on machine learning like 
{\tt deepCR} \citep{Zhang_2020}, a deep-learning algorithm, may offer further improvements.

In contrast, the literature on methods of imputing missing data is sparse. Currently, the most common approach is the median replacement, which replaces a bad pixel with the median of its neighbours. Algorithms that apply median replacement include {\tt LACosmic}. Some other packages, such as {\tt astropy.convolution}, take an average of the surrounding pixels, weighted by Gaussian kernels. An alternative is a 1D linear interpolation. This approach is standard for the integral-field spectrographs GPI \citep{Macintosh_2008} and SPHERE \citep{Claudi_2008}. {\tt deepCR}, on the other hand, predicts the true pixel values by a trained neural network. However, none of these methods are statistically well-motivated, and they usually apply a fixed interpolation kernel to all images. In reality, however, the optimal kernel could vary from images to images. Moreover, in a continuous region of bad pixels or near the boundary of an image, most existing data imputation approaches either have their performance compromised or have to treat these regions as special cases. Only {\tt deepCR} can handle them naturally with minimal performance differences.

In this paper, we present \codename, a robust and flexible data imputation approach based on Gaussian Process Regression (GPR). GPR defines a family of interpolation kernels, within which we choose the optimal kernel for each image, using a training set that is extracted automatically from that image. We structure the paper as follows: Section \ref{sec:GPR} introduces GPR and the covariance function, explains how they lead to a parametric family of interpolation kernels and shows how this formalism handles additional missing data naturally. Section \ref{sec:Optimize} discusses our strategies of choosing the optimal kernel. Section \ref{sec:Implement} specifies the implementation of our algorithm. Section \ref{sec:example} shows four examples---a CCD image of a globular cluster, a spectroscopic CCD exposure, a raw CHARIS image, and PSF photometry and astrometry on a stellar image---to demonstrate the consistently good performance of our algorithm. Finally, Section \ref{sec:conclusion} summarizes our approach, discusses its limitations, and proposes directions for future work.

\section{An Interpolation Kernel from Gaussian Process Regression} \label{sec:GPR}

GPR is a statistical method of inferring the values of a function $f(\vb*{r})$ at locations $\vb*{r_j}$, given that we know its values at locations $\vb*{r_i}$ and the covariance function $K(\vb*{r},\vb*{r'})$ that gives the covariance between any two points $\vb*{r}$ and $\vb*{r'}$. The true values of the function are assumed to be drawn from a multivariate Gaussian distribution with zero mean and an infinite-dimensional covariance matrix given by the covariance function:
\begin{equation}
    \begin{bmatrix} f(\vb*{r_i}) \\ f(\vb*{r_j}) \end{bmatrix}
    \sim
    N \left[0,\begin{bmatrix} K(\vb*{r_i},\vb*{r_i}) & K(\vb*{r_j},\vb*{r_i}) \\
    K(\vb*{r_i},\vb*{r_j}) & K(\vb*{r_j},\vb*{r_j})
    \end{bmatrix}\right].
\end{equation}
In many cases, we assume that we have only imperfect measurements $g(\vb*{r_i})$ of the true function values $f(\vb*{r_i})$.  We assume the measurement errors to be Gaussian with zero mean and covariance matrix $C_{\rm data}$. Then, the function values $f(\vb*{r_j})$ at locations $\vb*{r_j}$ are related to the measured values $g(\vb*{r_i})$ at locations $\vb*{r_i}$ by
\begin{equation}
    \begin{bmatrix} g(\vb*{r_i}) \\ f(\vb*{r_j}) \end{bmatrix}
    \sim
    N \left[0,\begin{bmatrix} K(\vb*{r_i},\vb*{r_i})+C_{\rm data} & K(\vb*{r_j},\vb*{r_i}) \\
    K(\vb*{r_i},\vb*{r_j}) & K(\vb*{r_j},\vb*{r_j})
    \end{bmatrix}\right].
\end{equation}
The function $f(\vb*{r_j})$ then has the expectation value
\begin{equation}
\left< f(\vb*{r_j})\right>=K(\vb*{r_j},\vb*{r_i})\left(K(\vb*{r_i},\vb*{r_i})+C_{\rm data}\right)^{-1}g(\vb*{r_i}).
\label{eq:gpr_expectation}
\end{equation}

GPR has been implemented in code packages such as {\tt GPy} \citep{gpy2014} and {\tt George} \citep{George_2015} and is commonly used for data interpolation and model inference, but its applications in image processing are rare because of the enormous size of imaging data. In the present paper we apply GPR to the problem of imputing missing data in a two-dimensional image. We take the function $g$ to be the measured counts by an instrument subject to read noise, shot noise, and instrumental effects; the function $f$ to be the true intensity on the sky; $\vb*{r_i}$ to be the good pixels; and $\vb*{r_j}$ to be the bad pixels where the measured values $g$ are not available. The problem is then equivalent to finding $\left< f(\vb*{r_j})\right>$ from $g(\vb*{r_i})$ by Equation \eqref{eq:gpr_expectation} given a covariance function $K$.  The values $\left< f(\vb*{r_j})\right>$ approximate the missing data at pixels $\vb*{r_j}$ and can replace them in the original two-dimensional image.

\begin{figure*}
    \includegraphics[width=\linewidth]{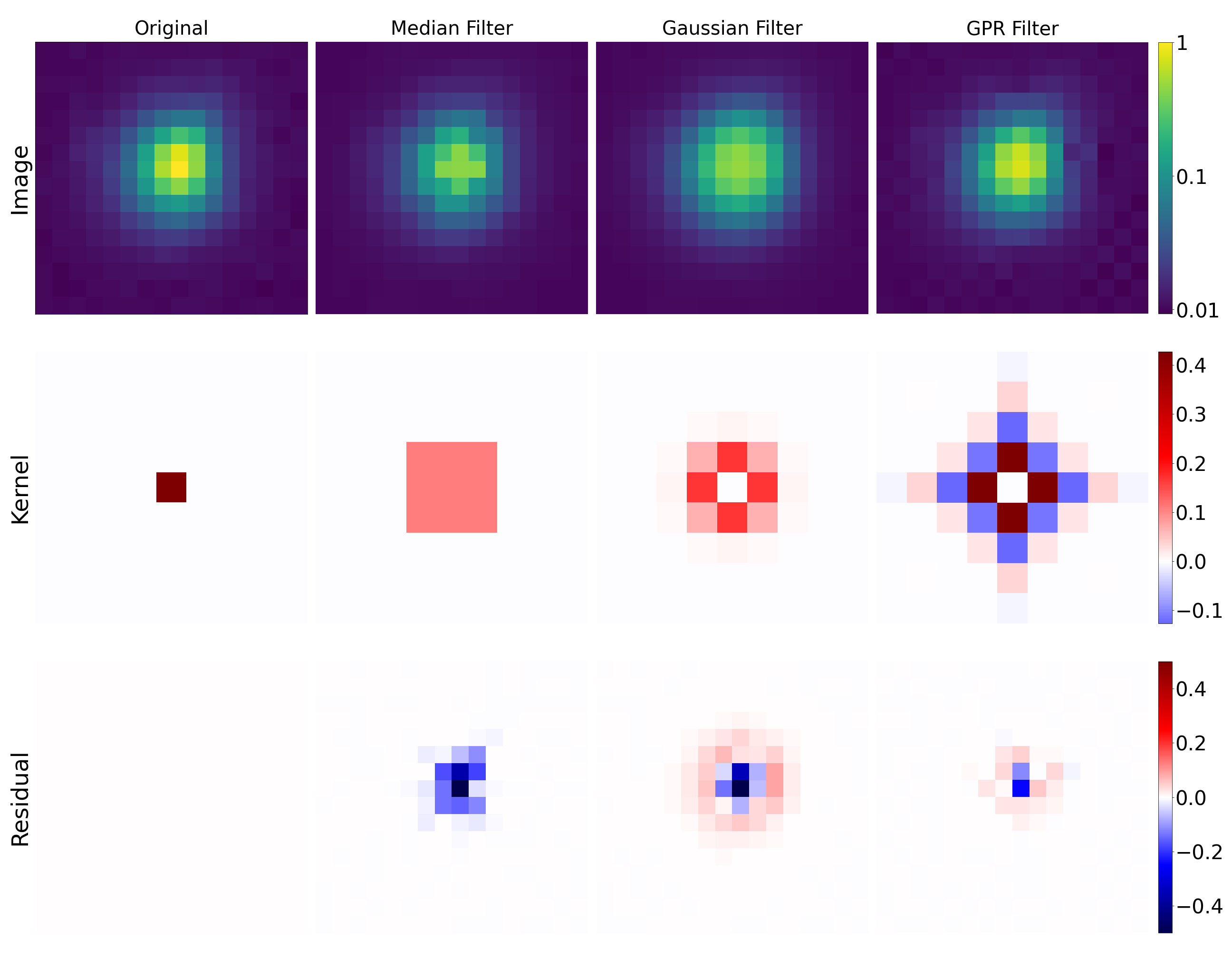}
    \caption{Part of an image (top left) convolved with three different kernels: a median filter (second column), a Gaussian kernel with zero weight in the central pixel and renormalized to unit area (third column), and a GPR kernel (right column). The GPR kernel best restores the original image. It is constructed from the squared exponential covariance function with $a=3.02$ and $h=0.72$ (which have been tuned as described in Section \ref{subsec: choose_parameters} using the full image). The Gaussian kernel has standard deviation $0.72$ to match the GPR kernel. The middle row shows the interpolating kernels themselves, and the bottom row shows the residuals of convolution. The original image is normalized by its brightest pixel. The kernels and residuals are plotted on a color scale such that white is zero, red is positive, and blue is negative.}
    \label{fig:GPR_filters}
\end{figure*}

Because of the ${O(n^3)}$ complexity of matrix inversion, using Equation \eqref{eq:gpr_expectation} on an entire image with several million pixels is not possible. We instead correct for one bad pixel at a time using only its neighbors. The resulting estimate of a bad pixel's value turns out to be a weighted sum of its neighbours' counts.  This may be thought of as an interpolation kernel applicable to the entire image. For example, if we use all pixels that are in the $9\times 9$ region centered at the bad pixel, then $f(\vb*{r_j})$ would be a scalar, $g(\vb*{r_i})$ an $80 \times 1$ vector, $K(\vb*{r_j},\vb*{r_i})$ a $1\times 80$ vector, and $\left(K(\vb*{r_i},\vb*{r_i})+C_{\rm data}\right)^{-1}$ an $80\times 80$ matrix. If $C_{\rm data}$ is diagonal with constant variance $\sigma_i \equiv \sigma$, and if $K(\vb*{r},\vb*{r'})$ depends only on the relative position $\vb*{r_i}-\vb*{r_j}$, then the $1\times 80$ vector $G \equiv K(\vb*{r_j},\vb*{r_i})\left(K(\vb*{r_i},\vb*{r_i})+C_{\rm data}\right)^{-1}$ would be the same for all $\vb*{r_j}$. Filling in zero for the missing central pixel (so that it is neglected when estimating its value) and reshaping to a $9\times9$ array, $G$ becomes an interpolation kernel. Figure \ref{fig:GPR_filters} compares the convolution of an image with a GPR kernel to that with a median filter and with a Gaussian kernel.

The residual of the convolution at a pixel can be thought of as the residual of image imputation if that pixel were the only bad pixel on the image. The convolution kernel that produces the least residual is thus the most suitable for image imputation. In our comparison, the GPR kernel best restores the original image because we have optimized it for image imputation 
using methods described in Section \ref{subsec: choose_parameters}. In Figure \ref{fig:GPR_filters} and all subsequent figures, the kernels and residuals are plotted on a color scale such that white is zero, red is positive, and blue is negative.

\subsection{The Covariance Function}

The choice of covariance function determines the smoothness and the extent of the kernel.  
In principle, a wide variety of covariance functions could be used, but we will focus on the squared exponential covariance function because of its simplicity and its good approximation to a typical instrumental PSF:
\begin{equation}
    K_{\rm SE}(\vb*{r},\vb*{r'})=a^2 \exp\left[-\frac{{|\vb*{r}-\vb*{r'}|}^2}{2 h^2}\right].
    \label{eq:squaredexp_cov}
\end{equation}
The parameter $a$ in Equation \eqref{eq:squaredexp_cov} describes the magnitude of the correlation between pixels; it should be compared to the data variance $\sigma^2$.  The kernel does not depend on $a$ and $\sigma$ individually but on the correlation-to-noise ratio $\alpha \equiv a/\sigma$, due to the fact that $K \propto a^2=\alpha^2\sigma^2$ and $C_{\rm data}\propto \sigma^2$, so $(K+C_{\rm data})^{-1} \propto \sigma^{-2}$, canceling $\sigma$ in Equation \eqref{eq:gpr_expectation}. Hence, in the rest of the paper we will set $\sigma=1$ (so that $C_{\rm data}$ is the identity matrix) and use $a$ to mean $a/\sigma$. The noiseless limit $a/\sigma \rightarrow \infty$ implies a constant, dark image, and a kernel independent of $a$. The other limit, $a/\sigma \rightarrow 0$, implies a noisy image where no useful information about a pixel's value can be extracted from its neighbors. The parameter $h$ gives the characteristic scale over which the image varies. Larger values of $h$ correspond to broader kernels.  The values of the hyperparameters $a$ and $h$, together with a particular choice of the covariance function, parameterize a family of GPR kernels.

\begin{figure*}
    \centering
    \includegraphics[width=\linewidth]{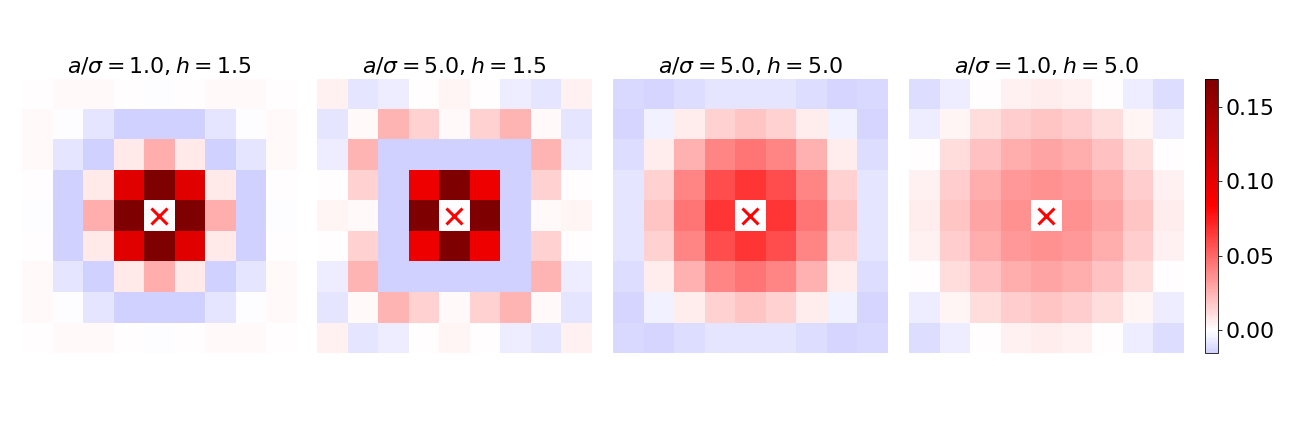}
    \caption{Example GPR kernels constructed using the squared exponential covariance function with different values for the hyper-parameters $a/ \sigma$ and $h$. An increased $a$ assigns larger weights and more significant negative weights. An increased $h$ causes the weights to change over a larger scale. All of the GPR kernels are zero at the center (the location of the bad pixel) and have ring-like structures due to the azimuthal symmetry of $K_{\rm SE}$.}
    \label{fig:kernels_varying_a_h}
\end{figure*}
Figure \ref{fig:kernels_varying_a_h} shows four examples of GPR kernels with different hyperparameters $a/ \sigma$ and $h$. As  $a/\sigma$ increases, GPR makes use of more pixels in the region and assigns larger weights to the adjacent pixels, allowing significant negative weights to be used. As $h$ increases, the absolute values of the weights decay more slowly with increasing distance; the weights change over a larger scale. All GPR kernels with a single bad pixel have ring-like structure because the value of the squared exponential covariance function depends only on the Euclidean distance to the central pixel, and they are zero at the center where the bad pixel is located. 

\subsection{Additional Missing Pixels in the Kernel}
\label{subsec: additional_missing_data}

\begin{figure*}
    \centering
    \includegraphics[width=\linewidth]{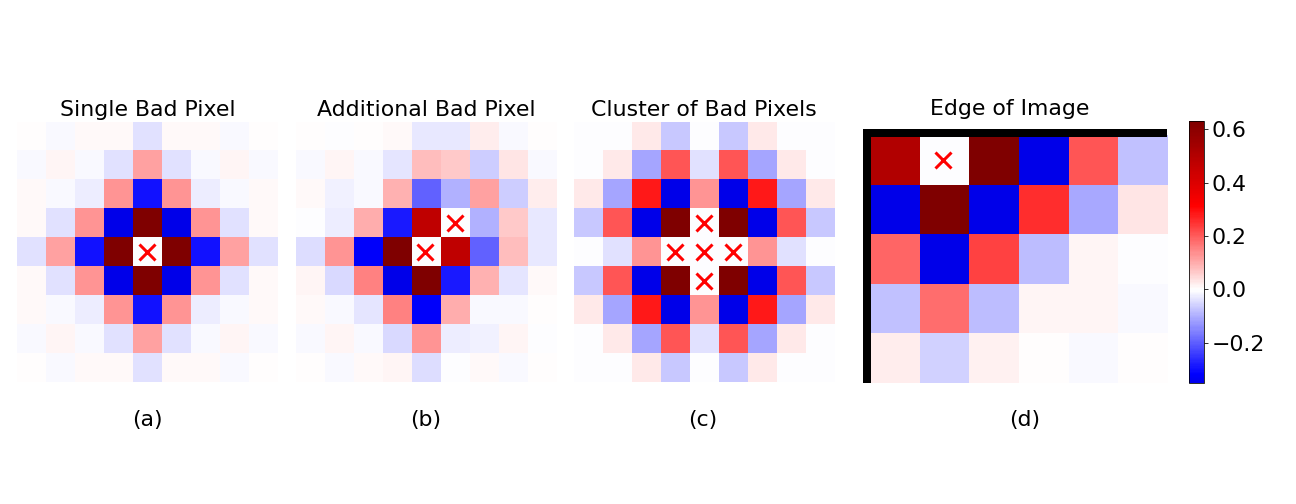}
    \caption{The re-weighting of GPR kernels due to additional missing data. All kernels are based on the squared exponential covariance function with hyperparameters $a/\sigma=10$ and $h=1$. Panel (a) shows the original kernel. Panel (b) has a bad pixel at $(1,1)$, so the pixels in that direction are assigned larger weights to compensate for the loss of information. Panel (c) has a cross of bad pixels at the center, so the weights that they originally had are redistributed to their nearest neighbours. Panel (d) shows a bad pixel at location $(1,0)$ relative to the top left corner of the image, so we are only able to construct a $5\times6$ kernel and the weights change accordingly.}
    \label{fig:GPR_reweighting}
\end{figure*}

The fact that each particular choice of $a$, $h$, and covariance function uniquely determines a convolution kernel has several advantages. One of them is that when there is a continuous region of bad pixels to be assigned zero weights, GPR is able to recompute the weights on all other pixels instead of simply re-normalizing them. To give zero weights to additional bad pixels near the bad pixel to be corrected, all we need to do is remove the corresponding entries in the covariance function and GPR automatically compensates for the loss of information by increasing or reducing the weights of their neighbours. Figure \ref{fig:GPR_reweighting} panels (a) through (c) show how additional bad pixels near the central bad pixel re-weight the entire GPR kernel and break its symmetry. 

 Another advantage of GPR is its straightforward handling of bad pixels on the edges of an image. Similar to the accommodation of additional bad pixels, we only need to remove all points that are outside of the image from the covariance function, and GPR will recompute the weights of the rest of the points. Figure \ref{fig:GPR_reweighting} panel (d) gives an example of a truncated GPR kernel due to a bad pixel on the top left corner of the image. Because the bad pixel is at $(1,0)$ relative to the corner, with a kernel width of $9$, we are only able to construct a $5\times 6$ kernel. GPR enables the weights to change accordingly to deal with the loss of information.

\section{Optimizing the GPR Kernel} \label{sec:Optimize}
Because $a$, $h$, and the covariance function parameterize a family of convolution kernels, we can optimize 
within the parameter space for each image to obtain a kernel that is the most suitable for repairing that image. In the high signal-to-noise ratio (SNR) limit, the optimization process can be viewed as modelling the instrumental effects on the structure of an image. For an image of distant stars, for example, it is roughly equivalent to finding the correlation function that best represents the instrumental PSF. 
The optimal kernel should not change as long as this PSF is stable across the image. Variations in the image itself, such as a spatially changing flux, do not affect the optimal kernel as long as the kernel is normalized. Hence, we optimize only once per image, and in general, one can optimize only once per instrument and per configuration.

\subsection{Choosing the Hyperparameters} \label{subsec: choose_parameters}

To choose the optimal hyperparameters $a$ and $h$, our strategy is to identify a set of well-measured pixels in the image to serve as the training set. Given $a$ and $h$, we can then convolve the entire image by the kernel defined by $a$ and $h$ and compute the subsequent change in value at each pixel in the training set. We find the $a$ and $h$ that minimize the mean absolute deviation of these residuals. 

The training set has to be large enough to be representative of the image, and it should focus on the pixels of scientific importance. Generally this means pixels with count rates appreciably higher than the background level. A kernel trained on low-SNR background pixels would simply learn to average a large set of surrounding pixels rather than identifying the correlation between pixels containing astronomically useful flux. A kernel optimized on a high-SNR training set would be good at correcting pixels of scientific importance even for low-SNR images (see Section \ref{subsec:low_SNR_example}). For a typical application, then, the training set should consist of pixels well above the background level but below saturation. Section \ref{subsec:implementation_training_set} describes \codename's choice of training set.

\subsection{Choosing the Optimal Covariance Function}

\begin{figure}
    \centering
    \includegraphics[width=\linewidth]{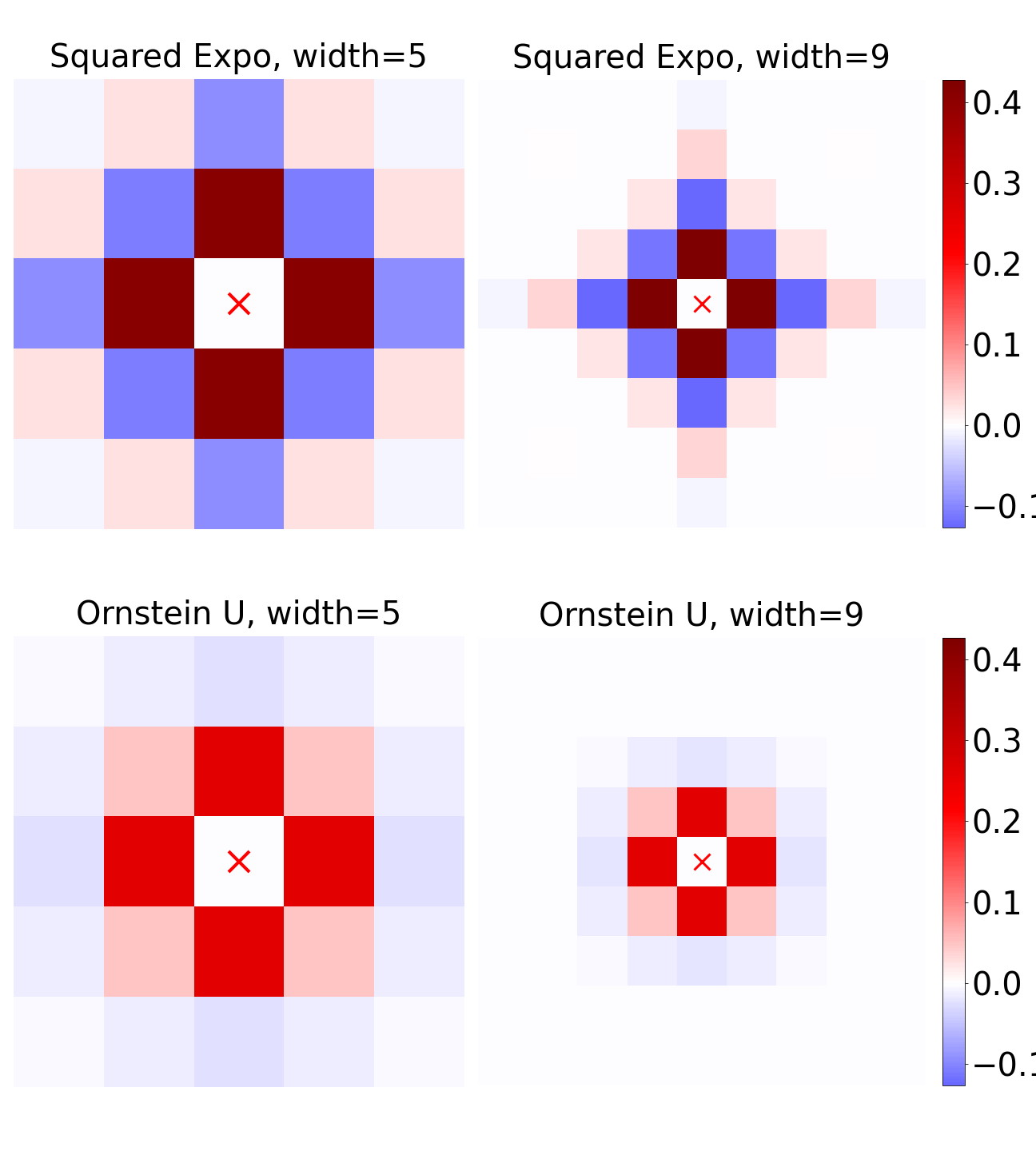}
    \caption{The optimized kernels for the star image in Figure \ref{fig:GPR_filters} using $K_{\rm SE}$ versus $K_{\rm OU}$ as the covariance function, and using kernel width $=5$ versus width $=9$. Due to the linear exponent in the covariance function, the $K_{\rm OU}$ kernels have smaller maximum weights, less prominent negative weights, and most of their weights are concentrated in the central $5 \times 5$ region no matter what kernel width we choose.}
    \label{fig:covariance_function_comparison}
\end{figure}

In principle, any covariance function can be used to generate GPR kernels, but only covariance functions that resemble the instrumental PSF would be useful. The squared exponential covariance function is therefore a reasonable choice. One notable alternative that we investigate here is the Ornstein-Uhlenbeck covariance function:
\begin{equation}
    K_{\rm OU}(\vb*{r},\vb*{r'})=a^2\exp\left[-\frac{|\vb*{r}-\vb*{r'}|}{h}\right]
    \label{eq:OUcov}
\end{equation}
Compared to the squared exponential covariance function, the Ornstein-Uhelenbeck covariance function is not quadratic but linear in Euclidean distance in the exponent. As a consequence, it tends to under-emphasize pixels that are the closest to the central bad pixel, which often contain the most important information.  Figure \ref{fig:covariance_function_comparison} compares the optimized kernels for the star image in Figure \ref{fig:GPR_filters} using the Squared Exponential covariance function (Equation \eqref{eq:squaredexp_cov}) with those using the Ornstein-Uhelenbeck covariance function (Equation \eqref{eq:OUcov}). The magnitude of the weights in the central region is significantly smaller for the $K_{\rm OU}$ kernels than for the $K_{\rm SE}$ kernels.

One other feature of a $K_{\rm OU}$ kernel is that its more slowly changing function values lead to smaller fluctuations in the weights and less prominent negative weights. As a consequence, most outer pixels have almost zero weights. In fact, for most values of $h$ smaller than the kernel width, almost all of the weight in a $K_{\rm OU}$ kernel is concentrated in the central $5 \times 5$ region. Hence, a $K_{\rm OU}$ kernel of width $>5$ is almost identical to a $K_{\rm OU}$ kernel of width $5$, which is not the case for $K_{\rm SE}$ kernels.

\begin{figure}
    \centering
    \includegraphics[width=\linewidth]{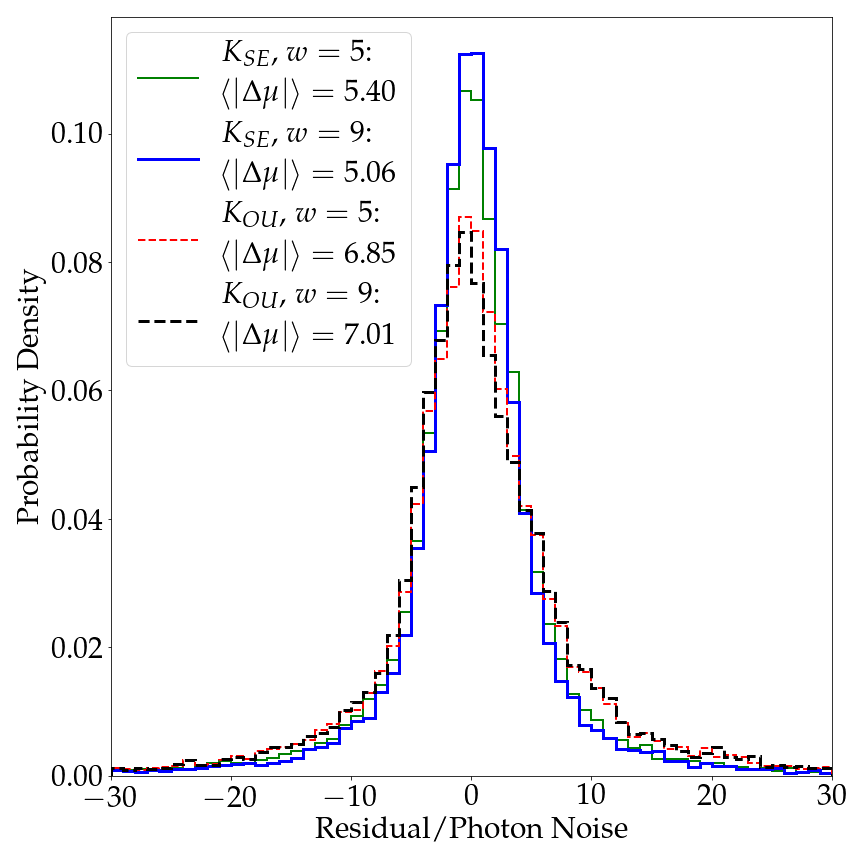}
    \caption{The residual (normalized by the $\rm shot \  \rm noise=\sqrt{\rm count/ \rm gain}$) of convolving the star image in Figure \ref{fig:GPR_filters} with the kernels in Figure \ref{fig:covariance_function_comparison}. Both $K_{\rm SE}$ kernels outperform the $K_{\rm OU}$ kernels, and increasing the width of a $K_{\rm SE}$ kernel from 5 to 9 further improves its performance.}
    \label{fig:covariance_function_residual}
\end{figure}

As a result of the two aforementioned features, a $K_{\rm OU}$ kernel is usually not as accurate and as flexible as a $K_{\rm SE}$ kernel. Figure \ref{fig:covariance_function_residual} shows the histogram of residuals (normalized by the $\rm shot \  \rm noise=\sqrt{\rm count/ \rm gain}$) from convolving the star image with the two types of kernels. In terms of the mean absolute deviations of the residuals, the $K_{\rm SE}$ kernel with width 9 performs better than the other kernels, while the $K_{\rm OU}$ kernel with width 9 has almost the same residual distribution as the $K_{\rm OU}$ kernel with width 5. 

\codename does have other covariance functions implemented, but we have found the squared exponential to produce the narrowest residual distribution.  We restrict further analysis in this paper to the squared exponential covariance function.

\subsection{A Stretched Kernel for 2D Spectroscopic Data}

\begin{figure}
    \centering
    \includegraphics[width=\linewidth]{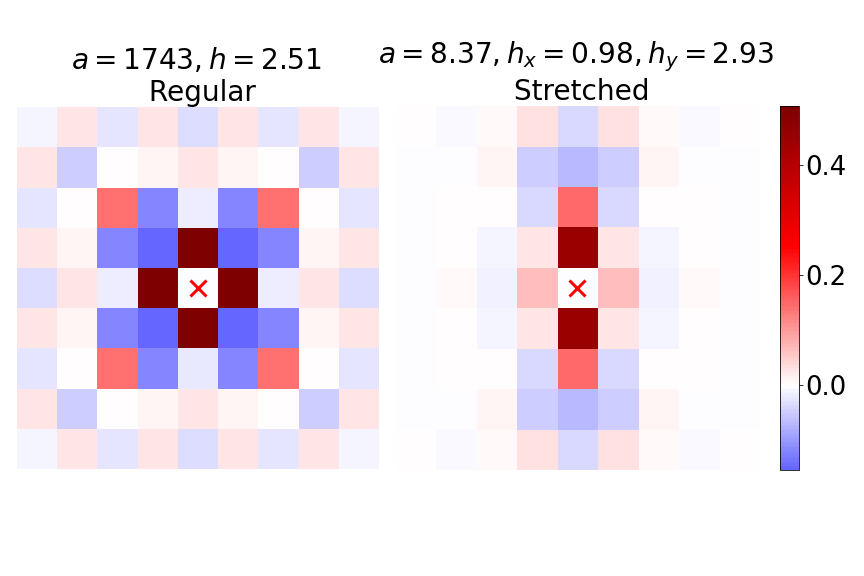}
    \caption{The isotropic versus the stretched kernel optimized for the spectroscopic arc lamp image in Section \ref{subsec: example_spectro}. The stretched kernel has greater weights in the $y$-direction because the slit is oriented vertically. Despite the different parameter values of the two kernels, they assign similar weights to the nearest pixels in the $y$-direction, which results in similar performance as shown in Figure \ref{fig:spectro_residual}.}
    \label{fig:Regular_VS_Stretched}
\end{figure}

So far, we have discussed GPR kernels that have weights depending only on the Euclidean distance $r$, but for anisotropic images, it is sometimes preferred to use a kernel that has directional dependence. Such a kernel can be particularly useful to the imputation of 2D spectroscopic data. For this purpose, we define the stretched squared exponential covariance function:
\begin{equation}
    K_{\rm sSE}(\vb*{r},\vb*{r'})=a^2 \exp\left[-\frac{1}{2}\left(\frac{(x-x')^2}{h_x^2}+\frac{(y-y')^2}{h_y^2}\right)\right]
\end{equation}
which produces kernels determined by three parameters: $a$, $h_x$, and $h_y$. We assume here that the preferred direction (e.g.~the dispersion direction) approximately follows one of the orthogonal axes of the detector, though it would be straightforward to adopt a different angle between the preferred direction of the image and the detector axes.  For isotropic images, we expect the optimization to give $h_x\approx h_y$. For vertical spectroscopic lines (horizontal dispersion) in a slit spectrograph, we expect that $h_x < h_y$ because the pixel values vary on a larger scale in the $y$-direction (along the slit). Figure \ref{fig:Regular_VS_Stretched} compares the isotropic and the stretched $K_{SE}$ kernel optimized for the spectroscopic arc lamp image in Section \ref{subsec: example_spectro}, and Figure \ref{fig:spectro_residual} shows that the stretched kernel produces smaller residuals in this example.

We encourage the user to exercise care with a stretched covariance function. The extra free parameter increases the risk of overfitting the hyperparameters. A stretched kernel can also smooth out a point source on the slit, especially if the hyperparameters are optimized on the sky background, a galaxy stretching the length of the slit, or a lamp flat.  Our training image in this case was an arc lamp for wavelength calibration; the best-fit hyperparameters would differ for a galaxy or a twilight spectrum.

\section{Implementation and Computational Cost} \label{sec:Implement}

We implement the GPR image imputation algorithm in the Python package \codename \citep{Zhang_2021}, which is available at \url{https://github.com/HengyueZ/astrofix}.
\codename accepts images with a bad pixel mask or images with bad pixels flagged as {\tt NaN}, and it fixes any given image in three steps:
\begin{enumerate}
    \item Determine the training set of pixels that \codename will attempt to reproduce.
    \item Find the optimal hyperparameters $a$ and $h$ (or $a$, $h_x$ and $h_y$) given the training set from Step 1.
    \item Fix the image by imputing data for the bad pixels using the optimized kernel from Step 2.
\end{enumerate}
We describe the details of each step in the subsections below. Appendix \ref{appendix:cost} details the computational cost of our approach.  The computation time is a function of the image size and the fraction of bad pixels. On a single processor, it is $\sim$10\,seconds for a $2000 \times 2000$ image with 0.5\% bad pixels, or a $4000 \times 4000$ image with 0.1\% bad pixels.

\subsection{Defining the Training Set}\label{subsec:implementation_training_set}
The set of pixels used to optimize the GPR should include as many pixels of scientific importance as possible, while avoiding pixels that are close to the background level, saturated, or themselves bad. By default, we include all good (unflagged) pixels that are below $1/5$ of the brightest pixel's value and $10$ median absolute deviations (MAD) above the median of the background. We take the median and the MAD of the image in lieu of the mean and standard deviation, because of the robustness and convenience of the median. The MAD is an underestimate of the standard deviation: for a Gaussian background distribution, $10$ MAD correspond to about $7\sigma$.

In many cases, the users can replace the default training set selection parameters ($1/5$ and $10$) with better ones by deriving the background distribution and the saturation level from the histogram of the image, but for high-SNR images with sufficient bright pixels, the quality of the training is largely insensitive to the parameters. The default training set usually works sufficiently well without requiring a highly accurate model of the background. For low-SNR images, one has to tune the parameters more carefully or use a high-SNR image taken by the same instrument (and under the same configuration) as the training set. 

The training set might include pixels that are themselves good but are close to bad pixels that can ruin the convolution. We remove such pixels during the training process described in the next subsection.

\subsection{Tuning the Hyperparameters}

In the high-SNR limit, the training process can be considered as modeling the structure that an instrument imparts to an image, such as the PSF. Therefore, an efficient strategy is to run the training only once for each instrument using a combined training set built from multiple images. However, the optimal kernel may still vary from image to image due to, e.g., variations in the seeing.
It is not computationally expensive to redo the training as appropriate for each image (see Appendix \ref{appendix:cost}). 

In the training process, we convolve the training set pixels with GPR kernels and vary the hyperparameters $a$ and $h$ (or $a$, $h_x$, and $h_y$) to minimize the mean absolute residual in the training set. We only need to compute the kernel once at fixed values for $a$ and $h$. The computational cost then scales as $N_{\rm train} \times w^2$, where $w$ is the width (in pixels) of the kernel and $N_{\rm train}$ is the number of pixels in the training set. We construct an image array of size $N_{\rm train} \times w^2$ to perform the convolution.

A problem naturally occurs in the limit of noiseless data, where the optimal value of the parameter $a$ (i.e.~$a/\sigma$ in our notation) would be infinite.  If $a\rightarrow \infty$ and $h$ is relatively large, evaluating Equation \eqref{eq:gpr_expectation} requires the inversion of an ill-conditioned (though not formally singular) matrix.  This leads to numerical instability in the constructed GPR kernel, and the optimization procedure can end up in a spurious minimum. 

\begin{figure*}
    \centering
    \includegraphics[width=\linewidth]{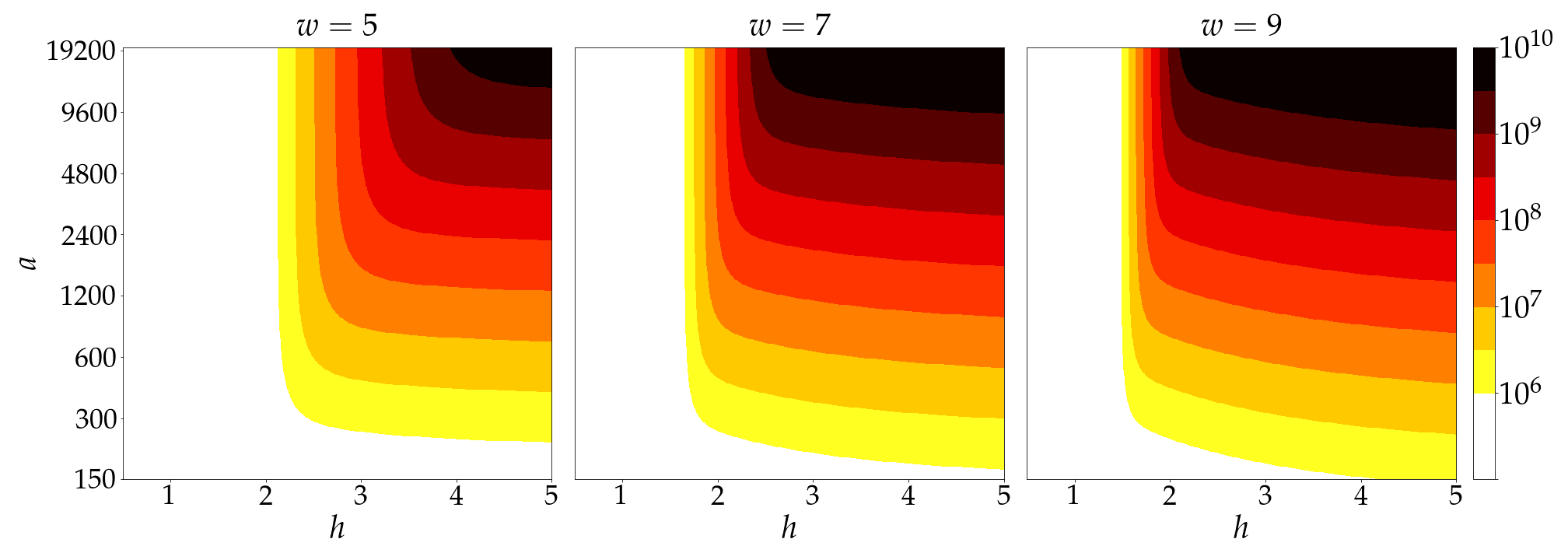}
    \caption{Dependence of the condition number on $a$ and $h$ for 3 different kernel widths. The condition number increases quickly with $h$ for small $h$ and increases quickly with $a$ for larger $h$. The contours indicate the upper bound on $a$ given a desired upper bound on the condition number.}
    \label{fig:cond_number}
\end{figure*}

We restrict $a$ to small values by multiplying the mean absolute residual with a penalty function:
\begin{equation}
    P=1+\exp\left[\frac{\left(a-\mu\right)}{\tau}\right]
\end{equation}
Here, $\mu$ can be interpreted as a "soft" upper bound on $a$. For $a\ll\mu$, $P\approx 1$; for $a>\mu$, $P$ increases exponentially. $\tau$ determines how much $a$ can penetrate the soft upper bound. The best values of $\mu$ and $\tau$ are determined by the largest condition number that the machine can handle in a matrix inversion without much loss of accuracy. Figure \ref{fig:cond_number} shows how the condition number depends on the hyperparameters $a$ and $h$ for 3 different kernel widths. A larger kernel width requires inverting a larger matrix, which leads to a larger condition number. The condition number increases quickly with $h$ for small $h$ and increases quickly with $a$ for larger $h$. The contours indicate the upper bound on $a$ given a desired upper bound on the condition number. For instance, if the maximum acceptable condition number is $10^8$, then in the worst case scenario (largest $w$ and $h$), $\mu=2400$ is roughly the upper bound on $a$. By default, we set $\mu=3000$ and $\tau=200$, which corresponds to a maximum condition number of about $6\times10^8$ for $w=9$ and $h=9$. The user can change the upper bound parameters according to their own needs and referring to Figure \ref{fig:cond_number}. 

Then, for a given image and a chosen training set, \codename's steps to tune the GPR hyperparameters are:
\begin{enumerate}
    \item Flag all bad pixels as NaN.
    \item Construct the image array for pixels in the training set, and remove rows that contain NaN values.
    \item Propose the values of $a$ and $h$ (or $a$, $h_x$, and $h_y$), and construct the corresponding kernel.
    \item Convolve the image with the kernel, restricting the convolution to pixels in the training set.
    \item Find the mean absolute residual of the convolved image and the original image for pixels in the training set.
    \item Minimize the mean absolute residual times the penalty function with respect to the parameters.
    \item Return the optimal parameter values and the residual distribution.
\end{enumerate}

Steps 3 through 5 are wrapped into one single function to be passed to {\tt scipy.optimize.minimize}, which is responsible for step 6. In step 3, we do not reweight the kernel in the presence of additional bad pixels because of its expense. We simply remove, in step 2, all affected pixels from the training set by checking each row of the image array for NaN values.
In step 6, we require that $a\geq 1$ because $a<1$ would imply that the correlation between pixels is smaller than the measurement noise. If $a$ is small, neighboring pixels would contribute little information about a missing data point and our effort would be meaningless. We also require that $1/2\leq h\leq w$. A characteristic scale larger than the kernel width would imply a spatially uniform image, while one much smaller than a pixel would again mean that a pixel's neighbors do not contain meaningful information to interpolate its value.  

The computational expense of the optimization process depends on the size of the training set, but it usually takes only a few seconds. For example, it took $\approx$6 seconds to train on the CHARIS spectrograph in Section \ref{subsec:example_CHARIS} with $\approx$439,000 
pixels in the training set. Appendix \ref{appendix:cost} includes a detailed analysis of the computational cost of \codename.

\subsection{Fixing the Image}
The optimal kernel found in the training process relies on the assumption that for every bad pixel, no other bad pixels or image edges lie within the $w \times w$ region of the kernel. In this case, correcting the image is equivalent to a simple convolution with the optimal kernel at the location of each bad pixel. However, if there are additional bad pixels or pixels off the edges of the image, we must set their weights to zero and recompute the entire kernel by removing the corresponding entries in the covariance function (as discussed in Section \ref{subsec: additional_missing_data}). When the bad pixel density is large, almost every bad pixel requires computing a new kernel, so the resulting computational cost scales linearly with the bad pixel density and can be expensive (see Appendix \ref{appendix:cost}). 

The optimal kernel that corresponds to an isolated bad pixel is usually used many times, so we pre-construct it before looping through all bad pixels. 
Because we correct only one bad pixel at a time and only a small fraction of the image overall, we use the weighted sum of the neighboring pixels rather than convolving the entire image. Finally, we replace the bad pixels with their imputed values only after exiting the loop. This avoids using imputed data as if they were actual measurements.

To sum up, \codename's steps to impute the missing data in an image are as follows:
\begin{enumerate}
    \item Pre-construct the optimal kernel for an isolated bad pixel.
    \item For each bad pixel:
    \subitem Check if there are additional bad pixels or image edges nearby.
    \subitem If not, use the optimal kernel. Otherwise, construct a kernel that has the largest possible size ($\leq w$ in each direction) and has weights reassigned according to the locations of additional missing pixels.
    \subitem Compute the sum of the neighboring pixels weighted by the kernel to find the bad pixel's imputed value. Store this value.
    \item Replace the bad pixels with the imputed values.
    \item Return the fixed image.
\end{enumerate}

\section{Examples and Results} \label{sec:example}
We tested \codename on a variety of images with real and artificial bad pixels.  In this section we give a few examples to demonstrate its consistently good performance.

\subsection{Artificial Bad Pixels on a CCD Image of a Globular Cluster} \label{subsec:example_cluster}
Images of star clusters often contain a high density of useful information. While approaches like PSF photometry can handle missing data appropriately (if a good model of the PSF is available), such an analysis is not always possible.  Aperture photometry, for example, cannot tolerate missing data.  We therefore test \codename on a real CCD exposure of the globular cluster NGC 104, taken by the SBIG 6303 0.4m telescope at the Las Cumbres Observatory (LCO). We randomly generated 62069 ($\sim$1\%) artificial bad pixels and imputed their values.

\begin{figure*}
    \centering
    \includegraphics[width=\linewidth]{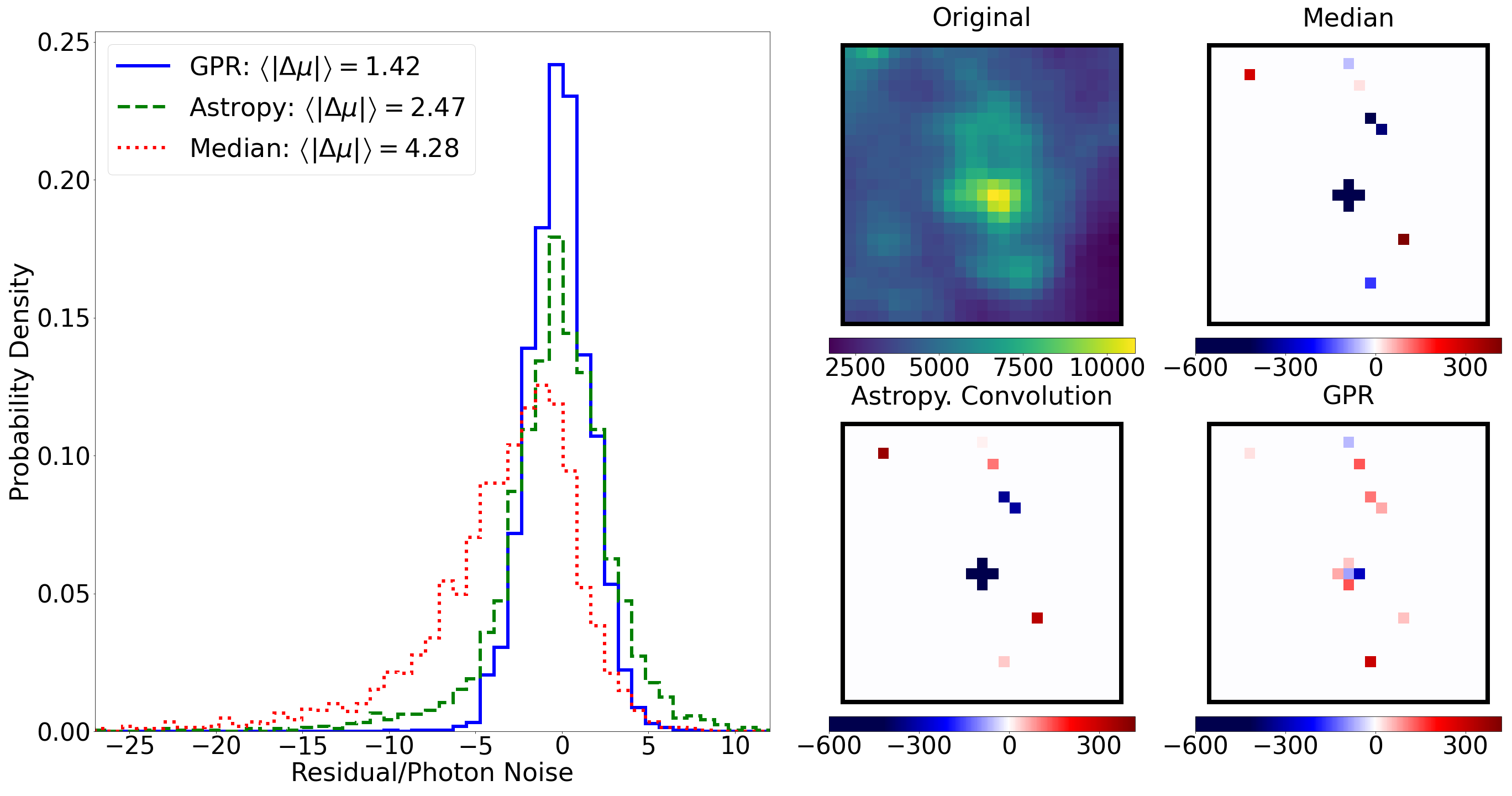}
    \caption{The residual ($\Delta\mu$) of correcting the cluster image by GPR, {\tt astropy.convolution}, and the $5 \times 5$ median filter used by {\tt LACosmic}. The histogram (left panel) includes the \changes{normalized} residual of the 2642 artificial bad pixels that have counts $10\sigma$ above the background. The right panels show the unnormlized residual in a $25 \times 25$ sample region. The good pixels are colored in white. Positive residuals are in red and negative residuals are in blue. The darkest blue points have values $<-600$: they are too bad to be differentiated by this color map. GPR outperforms {\tt astropy.convolution} by about a factor of 2 and median replacement by a factor of 3, and it performs especially well in dealing with continuous region of bad pixels.}
    \label{fig:cluster_residual}
\end{figure*}

The left panel of Figure \ref{fig:cluster_residual} compares the residual of image imputation $\Delta \mu$ by GPR to that by two standard approaches: {\tt astropy.convolution} (with a Gaussian kernel of standard deviation 1) and the $5 \times 5$ median filter used by {\tt LACosmic}. The residual is normalized by the \changes{shot noise}. Because most bad pixels that we have randomly selected were near the background level, we only include in the histogram the 2642 bad pixels that originally had counts 10$\sigma$ above the background.  This tests \codename's ability to restore useful information. The right panels of Figure \ref{fig:cluster_residual} show the unnormalized residuals in a $25 \times 25$ region at the center of the cluster. GPR outperforms the other approaches, especially in dealing with continuous regions of bad pixels. In terms of the outlier-resistant mean absolute deviation, GPR outperforms {\tt astropy}'s interpolation by about a factor of 2, and median replacement by more than a factor of 3. 

\subsection{Raw CHARIS Imaging Data} \label{subsec:example_CHARIS}

Infrared detectors are more specialized than CCDs, and often have a higher incidence of bad pixels. The CHARIS integral-field spectrograph (IFS), for example, uses a Hawaii2-RG array \citep{Peters+Groff+Kasdin+etal_2012}, and has a bad pixel fraction of about 0.5\% \citep{Brandt+Rizzo+Groff+etal_2017}. Similar instruments include the GPI \citep{Macintosh+Graham+Ingraham+etal_2014} and SPHERE \citep{Beuzit+Vigan+Mouillet+eatl_2019} IFSs; these have comparable bad pixel rates.  The data processing pipelines for GPI \citep{Perrin+Maire+Ingraham+etal_2014} and SPHERE both use box extraction to reconstruct the spectra of each lenslet. This approach cannot accommodate missing data, so both pipelines interpolate to fill in bad pixel values.

\begin{figure*}
    \centering
    \includegraphics[width=\linewidth]{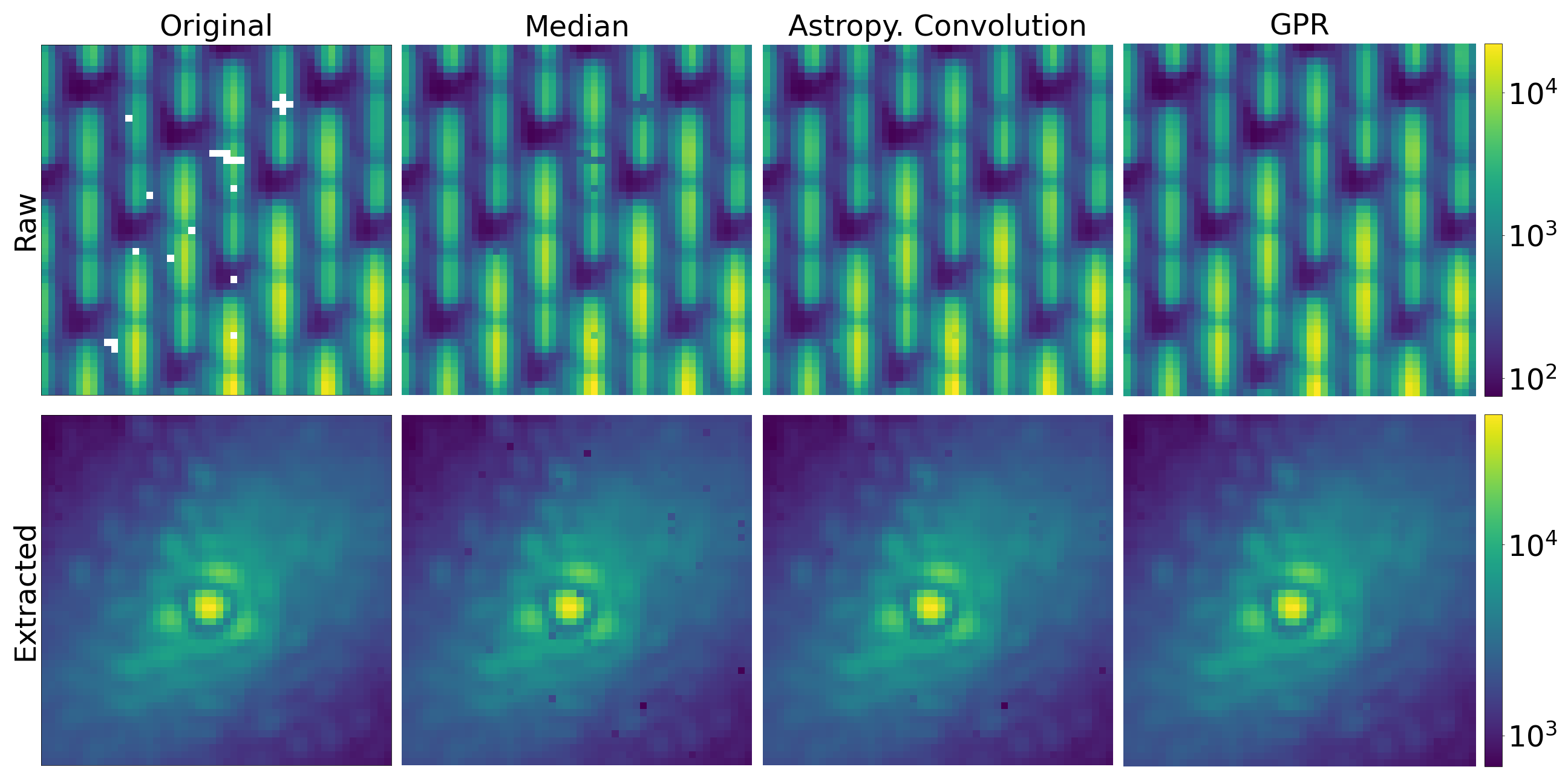}
    \caption{The top \changes{left} panel shows \changes{a $50 \times 50$ region in a raw CHARIS read with bad pixels, and the other top panels show the corrected images} by GPR, the $5 \times 5$ median filter, and {\tt astropy.convolution} ($\sigma=1$). The bottom panels show \changes{a $50 \times 50$ monochromatic slice through} the extracted data cubes. \changes{Then, with the CHARIS plate scale of 16.4 mas/lenslet \citep{Brandt+Rizzo+Groff+etal_2017}, each bottom panel shows the same $0.82^{\prime \prime}\times 0.82^{\prime \prime}$ field at $\lambda =1.80$ $\rm \mu m$, but extracted from different corrected images}. The original data cube comes from the standard CHARIS data reduction pipeline \citep{Brandt+Rizzo+Groff+etal_2017} which ignores bad pixels. GPR clearly outperforms both {\tt astropy} and the median filter and performs comparably well to the standard pipeline.}
    \label{fig:CHARIS_test}
\end{figure*}

We test \codename on a raw CHARIS image with 26338 intrinsic bad pixels flagged by values of zero. There were 438996 pixels in the training set, and the total computational time was 14.96 seconds. The top panels of Figure \ref{fig:CHARIS_test} show a slice of the raw image as well as the corrected images by GPR, by $5 \times 5$ median replacement, and by {\tt astropy.convolution} ($\sigma =1$). The bottom panels show a slice through the corresponding extracted data cubes. The original data cube comes from the standard CHARIS data reduction pipeline \citep{Brandt+Rizzo+Groff+etal_2017} which ignores bad pixels. GPR clearly outperforms both {\tt astropy} and the median filter and performs comparably well to the standard pipeline. When the data processing pipeline cannot ignore bad pixels (e.g. box extraction for GPI and SPHERE), \codename would become a useful interpolation method.

\begin{figure*}
    \centering
    \includegraphics[width=\linewidth]{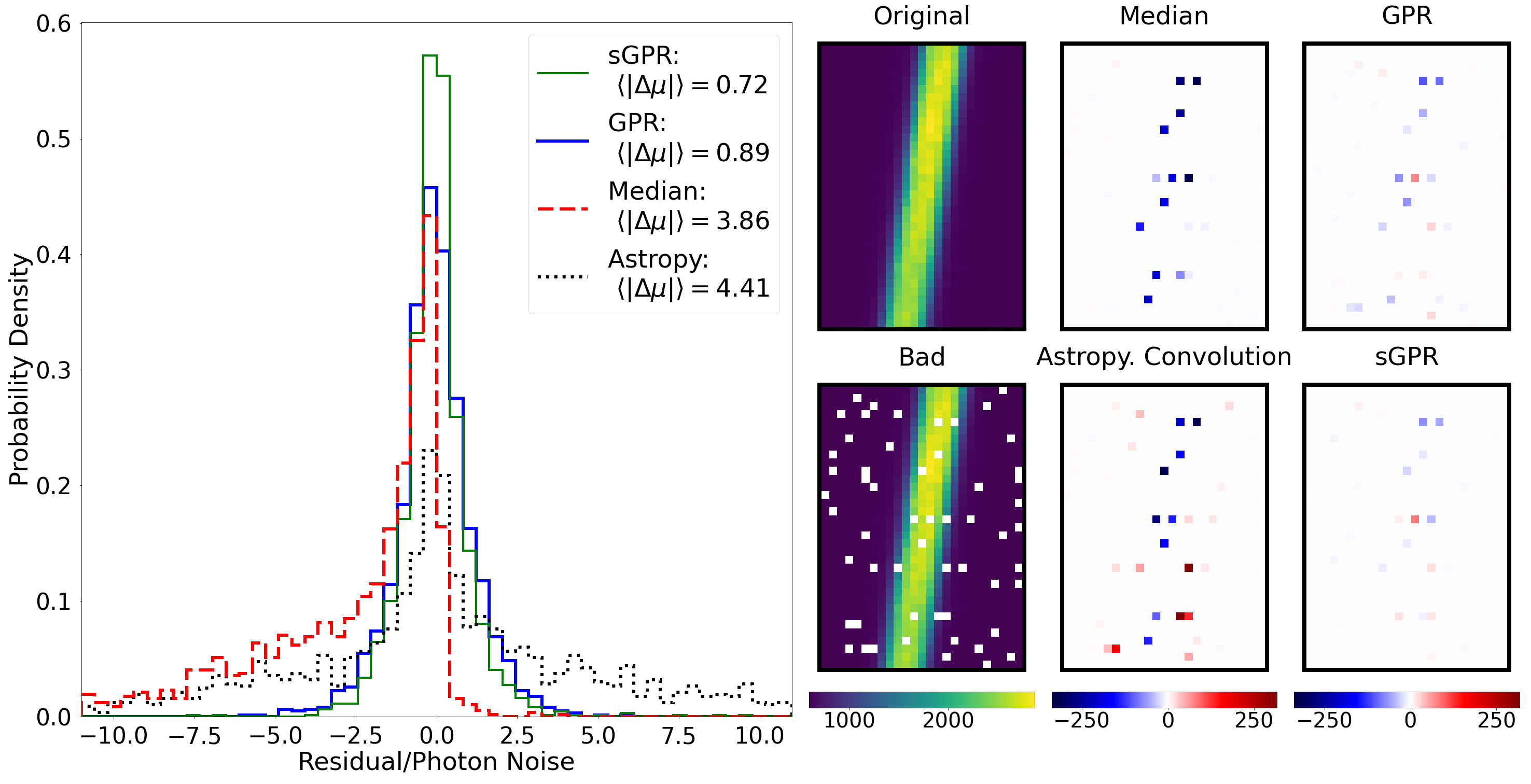}
    \caption{The residual ($\Delta \mu$) of correcting the 2D arc lamp spectrum by GPR, sGPR, {\tt astropy.convolution} ($\sigma =1$), and the $5 \times 5$ median filter. The histogram (left panel) includes the \changes{normalized} residual of the 1524 artificial bad pixels that had counts 10$\sigma$ above the background. The right panels show the unnormalized residual in a $35\times 25$ sample region. In this example, both GPR and sGPR outperform other approaches by a factor $>4$, and sGPR makes a small improvement over the conventional GPR.}
    \label{fig:spectro_residual}
\end{figure*}

\subsection{Artificial Bad Pixels on 2D Spectroscopic CCD Data} \label{subsec: example_spectro}
The standard GPR approach does not require any special treatment to adapt well to spectroscopic data, but we can apply a stretched GPR kernel to further improve its performance. In a real spectroscopic CCD exposure taken by the FLOYDS spectrograph \citep{Sand_2011} at LCO, we randomly generated 53537 ($\sim$5\%) artificial bad pixels, of which 1524 pixels had counts $10\sigma$ above the background. We applied both GPR and stretched GPR (sGPR) to impute the missing data, and their residuals \changes{(normalized by the shot noise)} are compared in the left panel of Figure \ref{fig:spectro_residual} with the residuals from {\tt astropy.convolution} ($\sigma =1$) and the $5 \times 5$ median filter. The right panels of Figure \ref{fig:spectro_residual} show the unnormalized residuals in a $35 \times 25$ region. Both GPR and sGPR outperform other approaches by a factor $>4$, and sGPR offers a small improvement over GPR. 

The data in this case are from an arc lamp and show little structure perpendicular to the dispersion direction. The results would differ for spectroscopy of a point source. Depending on the strength of spectral features, the image in the dispersion direction might appear smoother than in the slit direction, the reverse of this case.  For this reason, we urge caution in the use of sGPR with \codename.

\subsection{PSF Photometry and Astrometry on Low-SNR Stellar Images with Bad Pixels} \label{subsec:low_SNR_example}

The previous examples demonstrate \codename's ability to repair high-SNR images. In the low-SNR limit, however, the residual minimization process of \codename favors a uniform kernel with $a\rightarrow 0$ to smooth out the background noise. This reduces \codename to a simple average, which is suboptimal for restoring the useful information in a pixel. Therefore, we recommend that the user restrict the training to the brighter pixels in the low-SNR image or train on a high-SNR image taken by the same instrument (and under the same configuration) and apply the optimized kernel to the low-SNR image. Because such a kernel is trained on strongly correlated neighbouring pixels, it would be non-uniform and could produce a wider residual distribution for low-SNR bad pixels. However, the optimized kernel would perform substantially better than the uniform kernel in repairing pixels of practical use, such as pixels containing the PSF of a star.

As an example, we compare \codename and {\tt astropy.convolution} in terms of PSF photometry and astrometry on the truncated stellar image in Figure \ref{fig:GPR_filters}. We generate artificial bad pixels in its PSF and add white Gaussian noise to lower its SNR. We then apply both \codename and {\tt astropy.convolution} to repair the PSF at $\rm SNR=5,6,7,...,40$. We train the GPR kernel on the rest of the image with high SNR and get $a=3.02, h=0.72$. For {\tt astropy.convolution}, we apply a Gaussian kernel of $\sigma=0.95$ to match the $\sigma$ of the fitted PSF. Finally, we perform PSF photometry and astrometry on the fixed images and compute the deviations of the fitted centroids from the true centroid position ($\Delta x$ and $\Delta y$, in the unit of pixels) and the fractional difference between the fitted flux and the true flux ($\Delta F/F$). For both photometry and astrometry, we assume Gaussian PSFs and use {\tt IterativelySubtractedPSFPhotometry} and {\tt IRAFStarFinder} in the Python package {\tt Photutils} with the default parameters. We repeat the procedure for 1000 random noise samples at each SNR level.

\begin{figure*}
    \centering
    \includegraphics[width=\linewidth]{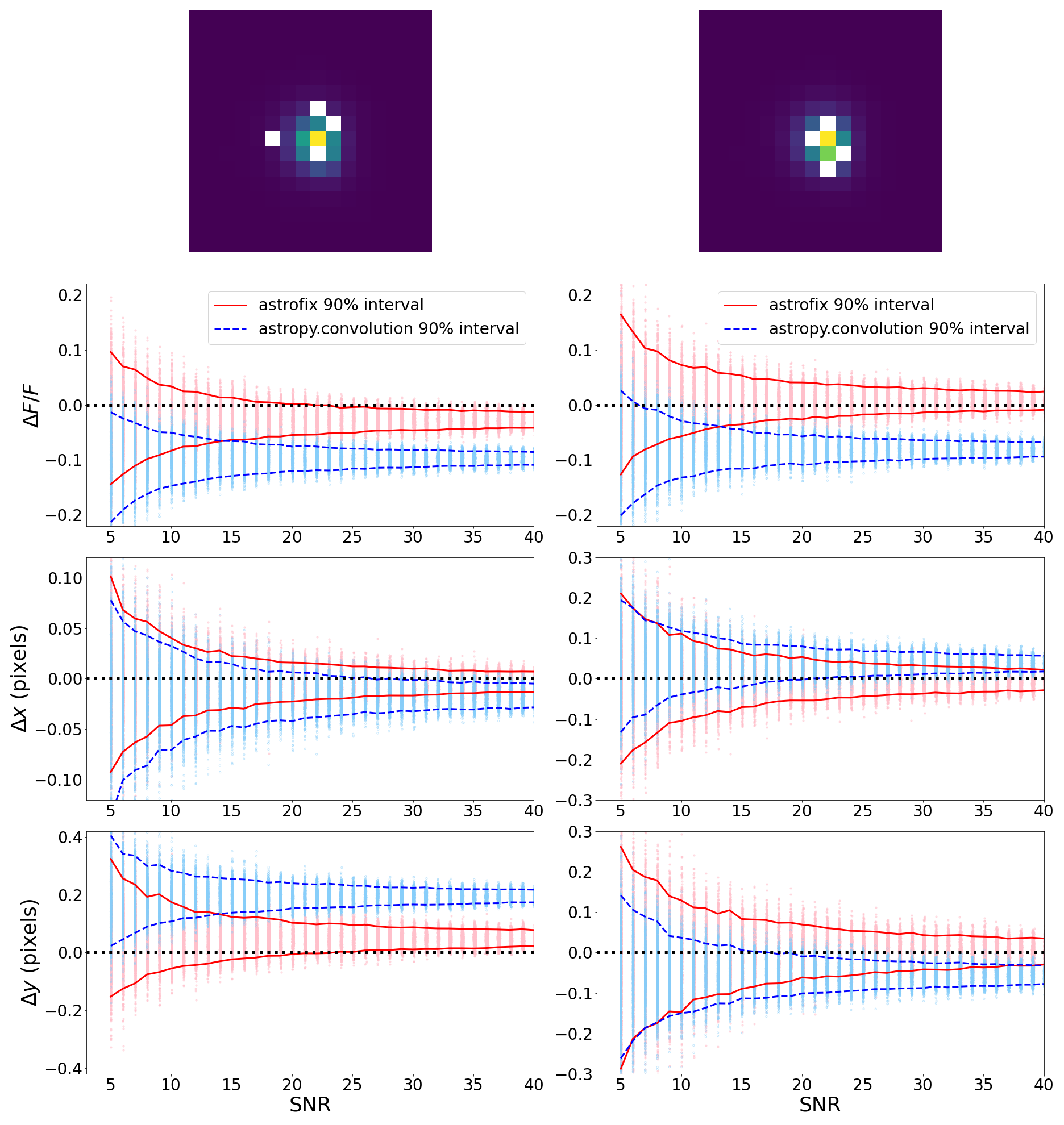}
    \caption{Performance comparison between \codename and {\tt astropy.convolution} in the low-SNR limit in terms of PSF photometry and astrometry. The bad pixels (white in the top row) in the PSF of the truncated stellar image in Figure \ref{fig:GPR_filters} are fixed by \codename and {\tt astropy.convolution} separately, and the difference between the PSF photometry and astrometry results on the fixed images and the true results are presented in the next three rows as a function of SNR. The GPR kernel was trained on the rest of the image with high SNR. The Gaussian kernel has $\sigma=0.95$ to match the $\sigma$ of the PSF. 1000 different noise samples are generated at each SNR level, and the thick lines enclose the 90\% intervals. For both bad pixel configurations, \codename produces much smaller systematic errors on all three fitted parameters, showing that the optimized GPR kernel is the more accurate model of the PSF, although, at very small SNRs, it generally produces a slightly larger spread (less precision) in the fitted parameters. This is because the GPR was trained to model the PSF instead of averaging out noise.}
    \label{fig:PSF_Photometry}
\end{figure*}

Figure \ref{fig:PSF_Photometry} shows the results for two different sets of bad pixels. \codename produces much smaller systematic errors on all three fitted parameters, showing that the optimized GPR kernel is the more accurate model of the PSF, although, at very small SNRs, it generally produces a slightly larger spread in the fitted parameters. In other words, we trade precision for accuracy by training the kernel on high-SNR pixels, putting greater weights on modeling the PSF instead of averaging out the noise. This strategy maintains the scientific value of \codename even for low-SNR images, and a high-SNR training set is almost always available from the brighter regions on the same image or other images taken by the same instrument.

\changes{A fully uniform background would remove the bias seen in Figure \ref{fig:PSF_Photometry}.  In this case, each neighboring pixel would contain the same information as the bad pixel, and the optimal approach would be a simple average of neighboring good pixels.  Our approach with \codename, however, will better retain small-scale structure atop a smoothly varying background.}

\section{Conclusions} \label{sec:conclusion}
We have presented \codename, an image imputation algorithm based on Gaussian Process Regression. The ability to optimize the interpolation kernel within a parametric family of kernels, combined with the simple and natural handling of clusters of bad pixels and image edges, enables \codename to perform several times better than median replacement and {\tt astropy.convolution}. It adapts well to various instruments and image types, including both imaging and spectroscopy.

As discussed in Section \ref{subsec:low_SNR_example}, the actual performance of \codename depends on the SNR of the image, and its optimization process requires a high-SNR training set. For a low-SNR image, the training set can only be obtained from other images taken by the same instrument, assuming that the instrument has negligible variations from images to images. The assumption is usually reliable, but it can sometimes fail due to changes in the seeing conditions, so it takes some extra effort to check these conditions before choosing a training set. The sGPR covariance function in \codename should be used with caution because it may smooth out a point source in the slit direction. The default parameters in \codename, such as the initial guess for the optimization and the cutoffs for the training set selection, were chosen carefully but have not been tested extensively on all image types. Nevertheless, they worked sufficiently well in all of our examples.

\changes{It is worth noting that \codename works best when there is sufficient structure on the scale of the kernel ($1/2 \leq h \leq w$). For a uniform background or a smooth, extended source varying only on a large scale $h\gg w$, the correlation-to-noise ratio would be extremely low within the size of the kernel, meaning that neighbouring pixels contribute almost no information to interpolate a pixel's value. In this case, \codename may not be as good as a simple average of neighbouring pixels. However, \codename will still outperform other methods in correcting bad pixels if the extended source does have structure on the kernel scale for \codename to learn, which is often the case. A better interpolation for each bad pixel then leads naturally to a better recovery of large-scale information.}

\acknowledgments{We thank the anonymous referee for an insightful and detailed report. This work made use of observations from the LCOGT network. It is our pleasure to thank Curtis McCully for sharing the LCO images used in the examples and for a helpful conversation. We also thank Tim Morton for his feedback on this manuscript and insightful suggestions on the \codename code. Finally, we would like to thank the entire UCSB Physics 240 class in the Spring 2020 quarter, where we discussed the very first ideas that lead to this work.

\software{{\tt Astropy} \citep{astropy:2013, astropy:2018}, {\tt Matplotlib} \citep{Matplotlib_2007}, {\tt NumPy} \citep{NumPy-Array2020}, {\tt Photutils} \citep{larry_bradley_2020}, {\tt SciPy} \citep{SciPy-NMeth2020}.}}

\appendix
\section{Computational Cost of astrofix} \label{appendix:cost}
\begin{figure}
    \centering
    \includegraphics[width=0.6\linewidth]{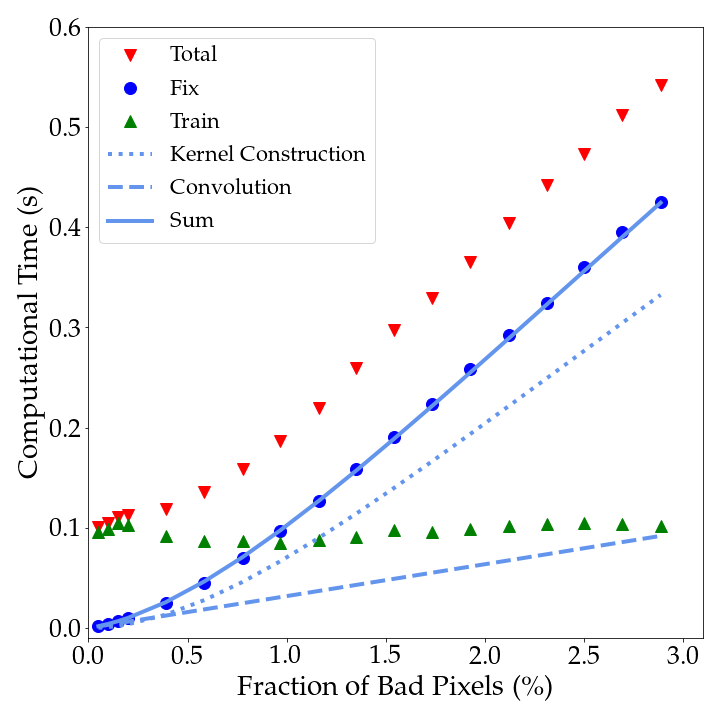}
    \caption{Computational cost of \codename in correcting a $200 \times 200$ region from the cluster image in Section \ref{subsec:example_cluster}. Each point is the median of 100 samples. The cost of fixing the image (solid blue line) depends linearly on $f$ for large $f$ and depends quadratically on $f$ for $f<1\%$. The cost of training stays roughly constant. Because of the complexity of matrix inversion, the cost of kernel construction (dotted blue line) is much greater than the cost of convolution (dashed blue line) for most values of $f$.
    }
    \label{fig:comp_time}
\end{figure}

Because \codename requires optimization and reconstruction of kernels, it has a greater computational cost than most existing image imputation approaches. Nevertheless, one can reduce the cost by using a smaller kernel width or by optimizing only once for each instrument without much loss in accuracy (see Figure \ref{fig:covariance_function_residual}). The computational cost also depends on the number of bad pixels in an image. In particular, the cost of training is largely independent of the bad pixel fraction, while the cost of fixing the image scales with the bad pixel fraction as follows:
\begin{equation} \label{eq:cost_scaling}
    C_{\mathrm{fix}}=AN_{\rm bad}+BN_{\rm bad}(1-(1-f)^{w^2-1})
\end{equation}
Here, the first term gives the total cost of convolving $N_{\rm bad}$ bad pixels with a kernel of width $w$, and the constant $A$ is the computational cost of $O(w^2)$ floating point operations. The second term gives the total cost of reconstructing the kernel, where $f=N_{\rm bad}/N_{\mathrm{tot}}$ is the bad pixel fraction and $(1-(1-f)^{w^2-1})$ is the probability that a bad pixel has at least one bad pixel neighbour in the surrounding $w\times w$ region. The constant $B$ is the average time of applying Equation \eqref{eq:gpr_expectation}, which consists of a matrix inversion of complexity $O(w^6)$ and a matrix-vector multiplication of complexity $O(w^4)$. As a consequence, $B\gg A$ and the cost of kernel construction is greater than the cost of convolution for large values of $f$. Assuming the bad pixels to be randomly positioned throughout the detector, the scaling of $C_{\rm fix}$ with $f$ depends on how $f$ compares to $w^{-2}$:
\begin{align}
    C_{\rm fix} \approx 
    \begin{cases}
    AfN_{\mathrm{tot}}+Bf^2(w^2-1)N_{\mathrm{tot}} & f \ll w^{-2} \\
    AfN_{\mathrm{tot}}+BfN_{\mathrm{tot}} & f \gg w^{-2}
    \end{cases}
\end{align}
In the limit of $f\gg w^{-2}$, $(1-f)^{w^2-1} \rightarrow 0$, and both terms in $C_{\rm fix}$ increase linearly with $f$. The cost per bad pixel stays constant because almost every bad pixel has at least one bad neighbour which requires reconstructing the kernel. 
In the limit of $f\ll w^{-2}$, $(1-f)^{w^2-1}\rightarrow 1-f(w^2-1)$, so $C_{\rm fix}$ depends quadratically on $f$. The cost per bad pixel becomes smaller because the kernel does not need to be recomputed for most bad pixels. 
The two limits are visible in Figure \ref{fig:comp_time}, which shows how the computational time of different components of \codename scales with the fraction of bad pixels on the cluster image in Section \ref{subsec:example_cluster}. Each point is the median of 100 samples. The solid blue line fits for $C_{\mathrm{fix}}$ using Equation \eqref{eq:cost_scaling}, while the dashed and dotted lines are the corresponding convolution cost and kernel construction cost. With $w=9$, only when $f<1\%$ does the scaling of the computational cost deviate from the linear relation of the large $f$ limit.  This linear scaling of the computational cost will set in at much lower values of $f$ if bad pixels tend to come in clusters (for example, due to cosmic rays).

For real images with real bad pixels, it usually takes $\sim$1 second on our Intel Core i5-8300H 2.30 GHz CPU to correct 2000 bad pixels, and the total time ranges from a few seconds to almost a minute. For example, the CHARIS data in Section \ref{subsec:example_CHARIS} took a total of 14.96 seconds, of which 8.56 seconds were spent on fixing 26338 bad pixels on the $2048\times 2048$ image, and 6.33 seconds were spent on the optimization process with 438996 pixels in the training set. The rest of the time went to defining the training set.

\bibliographystyle{apj_eprint}
\bibliography{refs}

\end{document}